\documentclass[rnote]{aa} 

\pdfoutput=1
\usepackage{graphicx}
\usepackage{txfonts}
\begin{document} 
\authorrunning{M.W. Blanco C\'ardenas et al.}
\titlerunning{Spectro-astrometry as a tool in the search of small sctructures
  in PNe}
   \title{CRIRES-VLT high-resolution spectro-astrometry as a tool in
     the search of small structures at the cores of Planetary Nebulae \thanks{Based on
       observations collected at the European Organisation for
       Astronomical Research in the Southern Hemisphere,
       Chile. Commissioning: 60.A-9051(A)}}

   \subtitle{}

   \author{M.W. Blanco C\'ardenas
          \inst{1}, 
          H.U. K\"{a}ufl
          \inst{2},
          M.A. Guerrero
          \inst{1}
          \and
          L.F. Miranda   
          \inst{3,4}
         \and
         A. Seifahrt
         \inst{5}   
          }

   \institute{Instituto de Astrof\'\i sica de Andaluc\'\i a (IAA-CSIC),
              Glorieta de la Astronom\'ia s/n, 18008 Spain\\
              \email{mblanco@iaa.es, mar@iaa.es}
         \and 
              European Southern Observatory (ESO), 
              Karl Schwarzschild Str. 2 D-85748 Garching bei M\"unchen, Germany\\
              \email{hukaufl@iaa.es}
         \and
             Universidad de Vigo, Departamento de F\'isica Aplicada,
             Facultad de Ciencias, Campus Lagoas-Marcosende s/n, 36310
             Vigo, Spain\\ 
             \email{lfm@iaa.es}
         \and
             Consejo Superior de Investigaciones Cient\'\i ficas (CSIC),
              C/ Serrano 117, E-28006, Madrid, Spain
         \and
            Department of Astronomy and Astrophysics, University of
            Chicago, Chicago, IL 60637, USA\\
            \email{seifahrt@oddjob.uchicago.edu}
}

   \date{Received September 15, 1996; accepted March 16, 1997}

 
  \abstract
   {
The onset of the asymmetry in planetary nebulae (PNe) occurs during
the short transition between the end of the asymptotic giant branch 
(AGB) phase and the beginning of the PN phase.  
Sources in this transition phase are compact and emit intensely in 
infrared wavelengths, making high spatial resolution observations in 
the infrared mandatory to investigate the shaping process of PNe.  
Interferometric VLTI IR observations have revealed compelling evidence of 
disks at the cores of PNe, but the limited sensitivity, strong observational 
constraints, and limited spatial coverage place severe limits on the 
universal use of this technique.  
Inspired by the successful detection of proto-planetary disks using 
spectro-astrometric observations, we apply here for the first time this 
technique to search for sub-arcsecond structures in PNe.  
Our exploratory study using CRIRES (CRyogenic
high-resolution Infra-Red EchelleSpectrograph) commissioning data of the proto-PN 
IRAS\,17516$-$2525 and the young PN SwSt\,1 has revealed small-sized 
structures after the spectro-astrometric analysis of the two sources.  
In IRAS\,17516$-$2525, the spectro-astrometric signal has a size of only 
12\,mas, as detected in the Br$\gamma$ line, whereas the structures found in SwSt\,1 have sizes of 230\,mas 
in the [Fe~{\sc iii}] line and 130\,mas in the Br$\gamma$ line. 
%
%
The spectroscopic observations required to perform spectro-astrometry 
of sources in the transition towards the PN phase are less time consuming
and much more sensitive than VLTI IR observations.  
The results presented here open a new window in the 
search of the small-sized collimating agents that shape the complex 
morphologies of extremely axisymmetric PNe.
}
   \keywords{techniques: spectroscopic 
-- high-resolution, ISM: jets and outflows 
-- planetary nebulae: general 
-- planetary nebulae: individual: IRAS\,17516$-$2525, SwSt\,1}
   \maketitle
%

\section{Planetary nebulae: shapes, shaping and observational challenges}

Planetary nebulae (PNe) emerge after the asymptotic giant branch (AGB) 
phase of low- and intermediate-mass stars ($0.8 M_\odot<M_i<8  M_\odot$).  
PNe display an exceptional variety of morphologies whose shaping involves 
dramatic changes in the otherwise round envelopes of their progenitors. 
The most axisymmetric PNe represent a challenge to the understanding of their 
formation, since they confront the canonical Generalized Interacting Stellar 
Wind model \citep[GISW, ][]{balick2002} .  
These PNe are believed to be sculpted by fast collimated outflows (jets), 
whose collimation and launch would be produced by strong magnetic fields 
\citep{GarciaS2005} and/or binary systems \citep{Soker98}.  
\cite{Marco2009} proposed the association of disks with
the binary evolution through a common-envelope phase.

As revealed by multiwavelength observations in the mid-IR and radio 
\citep{Sahai07}, the onset of the asymmetry in PNe occurs during the 
short transition between the end of the AGB phase and the beginning 
of the PN phase, when they are optically obscured.  
IR observations have revealed asymmetries in post-AGB sources and early 
proto-PNe such as bipolar lobes, and pinched waists and bright cores 
that appear as dark lanes at optical wavelengths 
\citep{Suarez2009,Suarez2011,Verhoelst2009,Lagadec2011,Blanco2013}.
Meanwhile, radio and mid-IR spectroscopic and interferometric observations 
have found strong evidences of small-sized dust and/or gaseous disks at the 
core of a few proto-PNe and young PNe.  
\citet{Bujarrabal2005} found a Keplerian disk of 900 mas in size at 
the {\bf centre} of AFGL\,915, whereas disks $\sim$30 mas have been detected 
by MIDI-VLTI in M\,2-9 and Mz\,3 \citep{Chesneau2007,Lykou2011} and 
a disk of 40 mas traced by water maser emission has been found in radio 
observations of the core of young PNe K\,3-35 
\citep{Miranda2001,Uscanga2008}.  

  \begin{figure*}
   \centering
   \includegraphics[width=15cm]{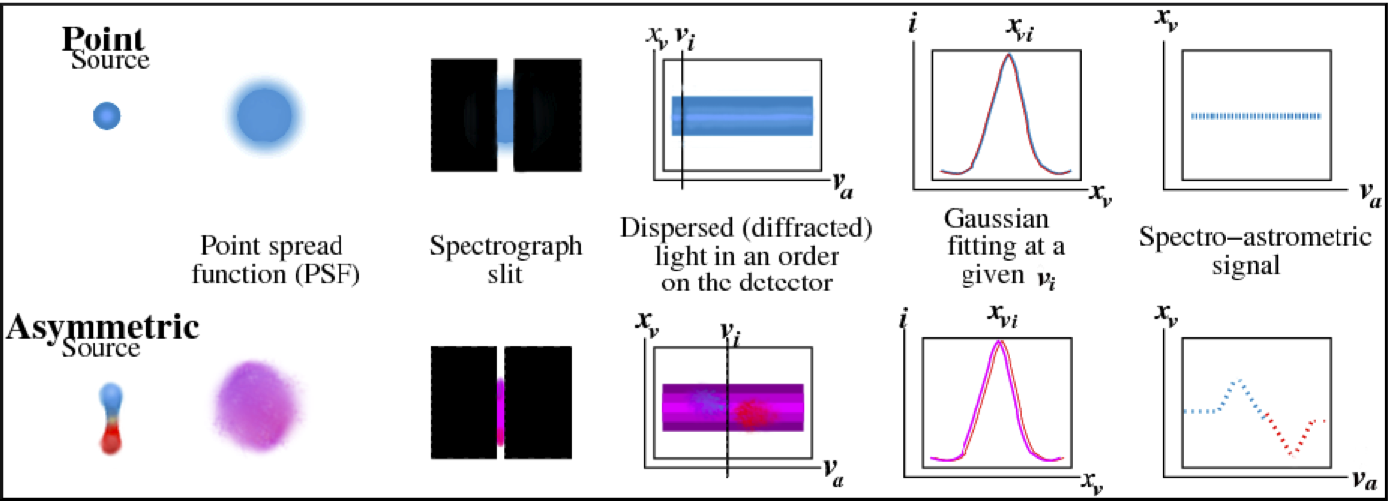}
     \caption{
Sketch of the SA technique \cite[adapted from][]{Whelan2008} showing the 
analysis of the long-slit spectra of a point source ({\it top}) and a 
compact bipolar PN ({\it bottom}).  
The major axis of the bipolar PN is assumed to be tilted with respect to 
the observer, with a lobe approaching (blue) and the other receding (red). 
The observation conditions smear out both sources into objects resembling 
unresolved point sources (magenta for the bipolar PN). 
The SA signal does not reveal any offset for the true point source, 
but those are unveiled for the bipolar PNe after the SA analysis. In
the latter, an antisymmetric SA signature arises because the velocity
shifts of the approaching and receding lobes are correlated with their
location at both sides of the adjacent central star continuum.
}
         \label{Fig1}
   \end{figure*}
%

These results illustrate the need for high-resolution IR and radio observations. 
MIDI-VLTI and AMBER-VLTI have achieved superb spatial resolutions, but 
these interferometric observations are challenging and time consuming.  
Furthermore, the interpretation of the few available baselines, which 
are not always suitably aligned with the orientation of the source on 
the sky, is strongly dependent on models.  
Finally, the sensitivity of these observations is limited.


An alternative to overcome the difficulties of VLTI observations
is the use of high-resolution spectro-astrometric (SA) studies.  
This observational technique has proven its efficiency in the search for 
close binaries \citep{Bailey98,Takami2001}, disks around B[e] stars 
\citep{Oudmaijer2008}, jets launched by brown dwarfs \citep{Whelan2005}, 
and protoplanetary disks \citep{Ponto08,Ponto11}.  
In the latter, Keplerian disks traced by the CO rovibrational fundamental 
band at 4.7\,$\mu$m with sizes in the range of 60--180\,mas \citep{Ponto08} 
have been detected using CRIRES-VLT \citep{Kaeufl2004}.
This instrumental setup overcomes previous attempts to use this 
technique with optical data that lacks the spatial resolution to
unveil small-sized nebular components \citep[e.g. HM Sagittae 
nova and the young PNe Vy 2-2, ][]{Solf84,Miranda91}. As compared to
spectro-astrometry at optical wavelengths the IR observations are less
affected by extinction. The resolved scales suggest rather compact
objects, hence being less sensitive to local extinction. This fact maybe a
decisive advantage of IR spectro-astrometry, when it comes to
reconstruct source geometries from the spectro-astrometric signatures.
Therefore, CRIRES high-resolution SA observations can open a new window 
in the search of compact structures in which high spatial resolution is 
crucial.

In this work we explore the SA capabilities of CRIRES for the detection and 
analysis of disk-like structures and small-scale asymmetrical features in 
objects in their transition to PNe. 
In Sect.~2 we describe the SA technique at high-resolution and the 
CRIRES commissioning observations used in this exploratory program.  
Preliminary results are described in Sect.~3 and a short summary is 
presented in Sect.~4

\section{CRIRES high-resolution spectro-astrometry}

The SA technique takes advantage of high-resolution spectral 
observations to recover spatial details at mas scales by adopting the 
stellar continuum adjacent to a spectral feature as an astrometric reference.
By using a low order adaptive optics module, CRIRES allows to push further the spatial 
resolution limit while rejecting artifacts which have been reported from 
the applications of the SA technique to seeing limited data. 
The SA technique is very attractive because it can be used to retrieve 
spatial information on scales that normally require interferometric 
observations, but using a single telescope with a standard instrumental 
setup.


The use of the SA technique is illustrated in Figure~\ref{Fig1} for a 
point source (a star) and a compact bipolar PN with its major axis 
tilted with respect to the observer. 
Both sources are blurred by the diffraction of the telescope, optical 
imperfections of the system, and residual seeing, and appear
unresolved according to the Rayleigh criterion, but 
the information of small-sized bipolar lobes is embedded within the 
dispersed spectrum.  
The SA analysis can trace the centroid of the signal along the spatial
axis ($x_v$) of the 2-dimensional spectrum as a function of the radial velocity 
($v_i$). The accuracy to trace this centroid will depend on the S/N,
which determines the PSF fitting precision.
The centroid at each velocity is determined by fitting a Gaussian 
function to the spatial intensity profile as such a function is a 
reasonable approximation to the point spread function (PSF) along 
the spatial direction of a dispersed spectrum \citep{stone89}.  
Once the centroid at a certain velocity ($x_{v_i}$) is computed for all 
velocities in the wavelength range ($v_a$) of interest, the variations 
of the spatial position of the centroid as a function of the velocity
can be traced without the dilution provoked by diffraction, optical 
imperfections, and seeing.  

The true spatial offset of a line may be more extended than measured
due to drag induced induced to its centroid by the continuum emission.
The drag depends on the relative intensities of line and continuum. 
In this work we have adopted \cite{Ponto08}'s formulation to derive 
and correct these effects on the SA signature, such that the corrected 
offset is computed as: 
\begin{equation}
\Delta X_{vi} = \Delta x_{vi}\times(1+F_c(v)/F_l(v)) 
\end{equation}
where $\Delta X_{vi}$ is the corrected offset, 
$\Delta x_{vi}$ is the offset measured by the Gaussian fitting, and 
$F_c$ and $F_l$ are the continuum and line fluxes, respectively.  
We will refer to $\Delta X_{vi}$ as the SA signature.

When the signal-to-noise ratio in the spectrum is sufficient, the 
original spatial structure will produce a SA signature.  
This signature will vary with the slit orientation: a physical structure 
tilted with respect to the plane of the sky and aligned along the slit will 
result in maximal SA signature, but minimal or no signature if oriented 
across the slit.  
A true point source will show a SA signature with no spatial offsets, 
independently of the orientation of the slit on the sky.  
On the contrary, a bipolar PN will show an antisymmetric SA signature 
because it has spectral signatures (the emission line from the approaching 
and receding lobes) correlated with spatial coordinates.  

The SA technique can reveal spatial structures with angular sizes down 
to small fractions of a pixel, on scales typically as small as 1/100th 
of the telescope diffraction pattern.  
We recall, however, that this technique does not produce direct images 
of a source;  only after a spatio-kinematical modeling of several slit 
positions, a given source geometry can be favored.  
In those cases when there is not enough information, the spatio-kinematical 
modeling may result in solutions that are not unique.

\section{Observations, data reduction, and analysis}


We have used CRIRES commissioning data to search for mas structures in 
sources in their transition towards the PN phase.  
CRIRES operates in the near IR, from 1 to 5 $\mu$m, offering the 
possibility to study ionized and molecular material, as well as 
the dust features already present in the near-IR L-band.  
CRIRES can reach a spectral resolution of $\sim$100,000 when the 
0$\farcs$2 slit width and the adaptive optic (AO) module are used.

The proto-PN IRAS\,17516$-$2525 and the young PN SwSt\,1 were observed 
on June 11, 2006, and June 5, 2006, respectively.  
The observations were taken with the AO loop closed and the 0$\farcs$2 
slit width. 
The slit was oriented along a PA of 0$\degr$, and two nodding positions 
were used, with exposure times of 120\,s.  
The observational template AutoNodOnSlit with a small nodding throw 
$\sim$10$\arcsec$, commonly used for compact sources, was used for 
IRAS\,17516$-$2525.  
The observational template Generic with a large nodding throw 
$>$25$\arcsec$, ideal for extended sources, was used for SwSt\,1.  
The data were registered by a 4$\times$1 mosaic of Raytheon 1024$\times$1024 
pixel InSb Aladin III detectors with pixel size of 27 $\mu$m.  
The detector readout mode was FowlerNsamp.  
The observations covered the spectral range 2.115--2.168 $\mu$m.

The data were reduced using standard IRAF procedures for long-slit 
spectra.  
Self developed IDL programs were used to complete the necessary 
corrections.  
The combined spectra were corrected for telluric absorptions using 
the spectro-photometric standard stars HR\,5985 for IRAS\,17516$-$2525 and 
HR\,7933 for SwSt\,1. 
The spatial resolution before the SA analysis, as derived from the 
full width at half maximum ($FWHM$) of the stellar continuum, was 
0\farcs44 for the Br$\gamma$ line of IRAS\,17516$-$2525, and 0\farcs70 
and 0\farcs66 for the Br$\gamma$ and [Fe~{\sc iii}] lines of SwSt\,1.  
We note that CRIRES was not optimally focused during this first 
commissioning run. 

The combined spectra used for SA analyses were corrected from detector 
misalignment at mas scales that might affect the signatures obtained by 
fitting the continuum position using a second order polynomial fit in
the case of SwSt\,1 and a third order polynomial fit for IRAS\,17516-2525.  
The centroid along the slit (spatial) direction was derived at each 
wavelength around lines of interest by fitting a Gaussian profile to 
the spatial intensity profile obtained in small apertures of 
2$\times$60 pixel$^2 \equiv$ 3 km\,s$^{-1}\times$ 5\farcs16. 
Additional IDL programs were written to perform the SA analysis.  
It is important to mention that, even though the observations here 
presented were not specifically tailored to this purpose, they have 
allowed us to develop the methodology and programing tools to apply 
the SA technique to search for mas structures in proto-PNe and young 
PNe.

%
%

\section{Spectro-astrometric preliminary results}

\begin{figure}
  \centering
  \includegraphics[width=0.7\columnwidth,bb=5 5 335 610]{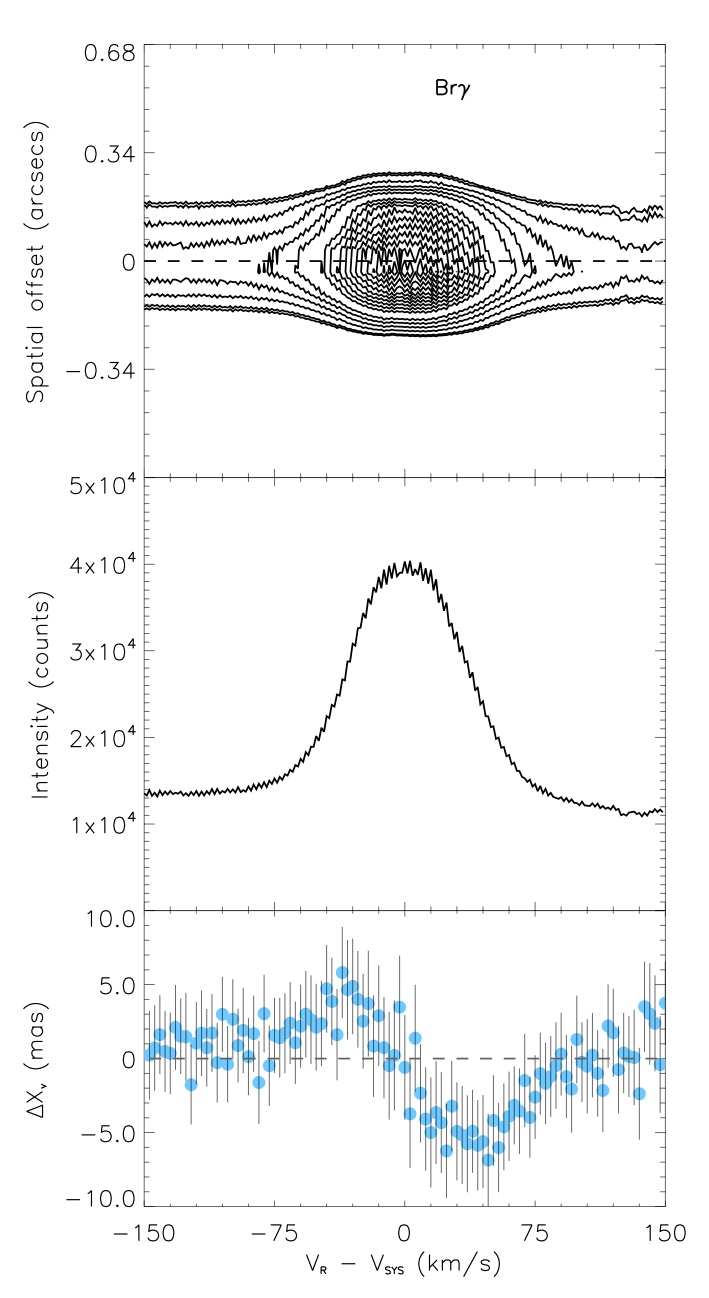}
  \caption{
PV map ({\it top}), line profile ({\it middle}), and SA signature 
({\it bottom}) of the Br$\gamma$ $\lambda$2.160 $\mu$m line of 
IRAS\,17516$-$2525. 
}
\label{Fig2}
\end{figure}

\begin{figure*}
  \centering
   \includegraphics[width=0.7\columnwidth,bb=0 0 311 575]{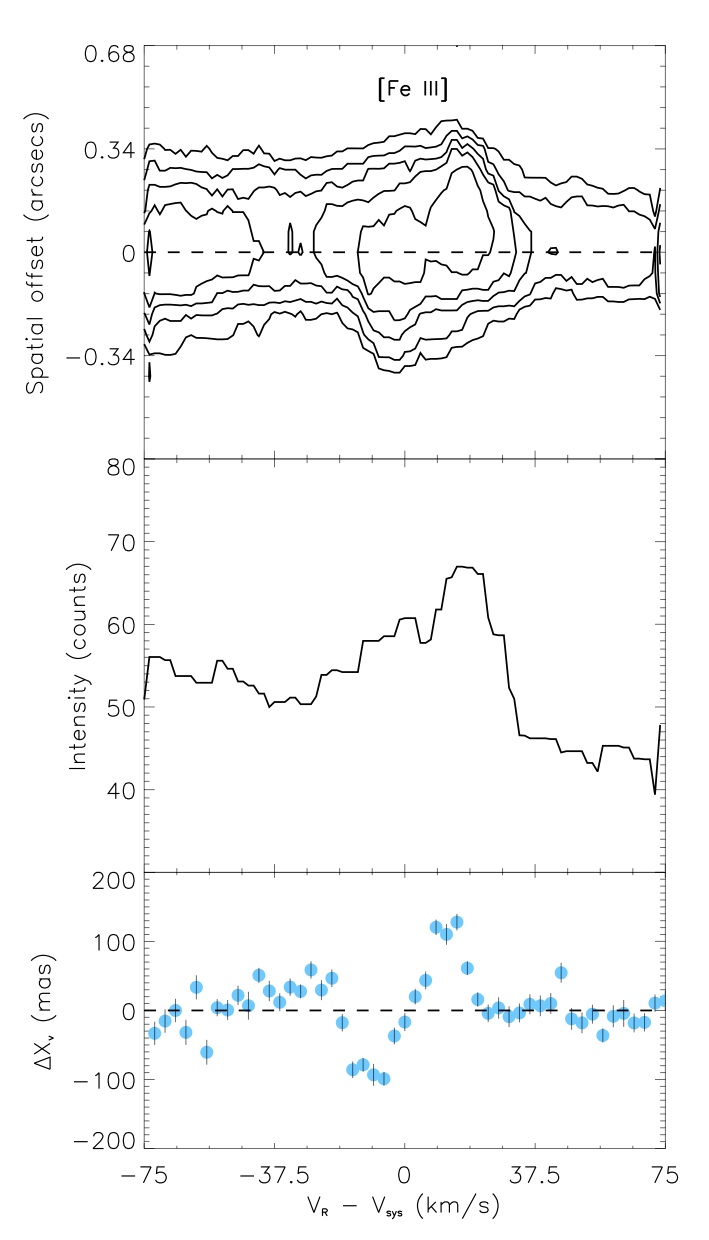}
   \includegraphics[width=0.7\columnwidth,bb=0 0 311 575]{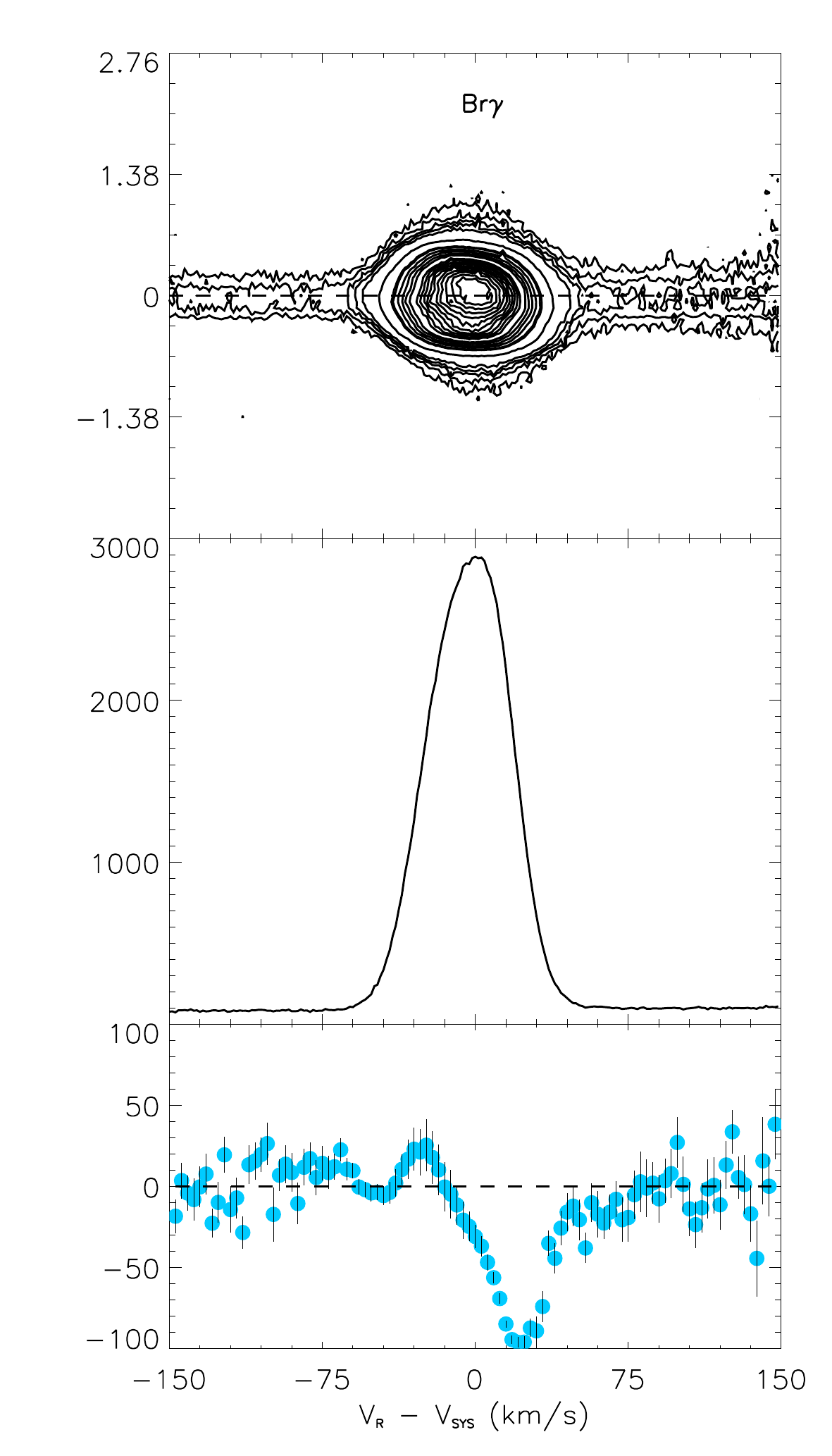}
  \caption{
PV map ({\it top}), line profile ({\it middle}), and SA signature 
({\it bottom}) of the [Fe\,{\sc iii}] $\lambda$2.14 $\mu$m ({\it left}) 
and Br$\gamma$ $\lambda$2.160 $\mu$m ({\it right}) lines of SwSt\,1.  
The SA signal for the two lines are basically reversed, suggesting 
that they trace different nebular components.  
}
\label{Fig3}%
\end{figure*}

%

IRAS\,17516$-$2525 has been classified as a proto-PN, although its morphology 
has not been revealed yet;  the source appears compact in near-IR images 
\cite[1$\arcsec$ in size,][]{RL09,RL2012} and it is not resolved in the 
mid-IR imaging catalogue of post-AGB stars presented by \citet{Lagadec2011}, 
despite the nearly-diffracted angular resolution of these images.
The spectroscopic study of S\'anchez-Contreras~et~al.\,(2008) found a 
P-Cygni profile in the H$\alpha$ line, but the Pa lines do not show 
this profile.  
We have selected the Br$\gamma \lambda$2.160 $\mu$m line detected 
in the CRIRES (Figure~\ref{Fig2}) spectrum to carry out the SA 
analysis. 
A preliminary inspection reveals that the line profile appears broad and 
single peaked.  
The SA analysis shows a compact antisymmetric position-velocity (PV) 
pattern extending $\sim$12 mas in size ($\Delta V \sim$80 km~s$^{-1}$). 
Although this SA signature is reminiscent of those interpreted by 
\citet{Ponto08} as protoplanetary disks, a disk rotating at this 
speed would require a gravitational potential unfeasible for a 
proto-PN.  
Small-sized bipolar lobes arising from the central regions 
of this source provide a more likely interpretation for 
this SA signature. 


SwSt\,1 is a compact PN with a C-rich central star.  
Spectroscopic studies have found P-Cygni profiles in the C~{\sc iii} 
$\lambda$4650 \AA\ and C~{\sc iv} $\lambda$5801 \AA\ lines, but these 
are not present in the H$\alpha$, H$\beta$, and H$\delta$ lines 
\citep{Marco2001}.  
The morphology of SwSt\,1 is not fully resolved in \emph{HST} images 
\citep{Marco2001}, but they suggest the presence of a compact, 
$\sim$2$\arcsec$ toroidal structure elongated in the east-west direction. 
In near IR K-band \citep{Likkel2006} and far-UV \citep{Sterling2005} 
observations, SwSt\,1 shows high iron abundances and inhomogeneities 
in the dust-to-gas ratio across the nebula.  

The PV plots of the [Fe\,{\sc iii}] $\lambda$2.145 $\mu$m and 
Br$\gamma$ $\lambda$2.160 $\mu$m (Figure~\ref{Fig3}) show certain 
asymmetry, particularly the [Fe\,{\sc iii}] line.  
The SA analysis reveals antisymmetric PV features $\sim$230 mas in size 
($\Delta V\approx$22 km~s$^{-1}$) in the [Fe\,{\sc iii}] line 
and $\sim$130 mas in size in the Br$\gamma$ line 
($\Delta V\approx$50 km~s$^{-1}$).  
Besides the different sizes, there is an obvious change in the shape and 
orientation of the signatures.  
This is reminiscent of sources in which distinct structural components 
with varying physical conditions are traced by different emission lines, 
such as the young PN Hb\,12 \citep{Welch99} and the proto-PN CRL\,2688 
\citep{Sahai98}.  
We can argue that in SwSt\,1 the [Fe~{\sc iii}] line (IP=30.7 eV) arises 
from a region in the nebula close to its central star, where the higher 
density and temperature of the material enhances its emissivity 
\citep{BP98,Zhang_etal2012}, whereas the location of the Br$\gamma$ line 
(IP=13.6 eV) emitting region is less constrained.

\section{Summary}

Using long-slit high-resolution spectra, the SA technique enables us 
to resolve small-sized structures in different astronomical objects.  
Inspired by these results, we have investigated the use of this 
technique in the search of structures at mas scales at the 
innermost regions of sources in the transition from the AGB to 
the PN phase.  
For this purpose, we have used CRIRES commissioning data of the proto-PN 
IRAS\,17516$-$2525 and the young PN SwSt\,1 to develop the methodology 
and the tools to perform such SA analyses.

Our SA analysis has been able to detect the presence of a 
small-sized structure, $\sim$12 mas in size, in IRAS\,17516$-$2525 
that can be interpreted as small bipolar lobes.  
As for SwSt\,1, the SA analysis shows two structures with different sizes, 
$\sim$130-230 mas, and inclinations in the [Fe\,{\sc iii}] and Br$\gamma$ 
lines.  
Tailored SA observations at different PAs with its corresponding 
calibrations, as well as a proper model fitting, are mandatory to 
derive a conclusive interpretation of the SA signatures.

Remarkably, the SA analysis based on CRIRES commissioning data has been 
able to detect structures which are comparable in size to those resolved 
by MIDI-VLTI in Mz\,3 and M\,2-9,  
and more compact than those detected in AFGL\,915 with radio-interferometric 
observations at Plateau de Bure.  
Compared to MIDI-VLTI, CRIRES SA observations are more sensitive, not 
restricted by fixed baselines, and the data interpretation is less model 
dependent.  
We conclude that the SA technique is very suitable for the search of 
small-sized disks and asymmetric structures in the short transition 
from the AGB to the PN phase.  



\begin{acknowledgements}
MWBC and MAG are supported by the Spanish MICINN (Ministerio de 
Ciencia e Innovaci\'on) grant AYA 2011-29754-C03-02, and LFM by 
grant AYA 2011-30228-C03-01, both co-funded with FEDER funds.  
MWBC would like to thank the EEBB-FPI for the grant for the short term stay 
in 2012, and thank the European Southern Observatory (ESO) Headquarters in 
Garching, Germany, for all the facilities provided during her stay.
We thank Dr.\ X.\ Fang for helpful discussion on the excitation of 
the [Fe~{\sc iii}] $\lambda$2.14 $\mu$m in PNe. 
\end{acknowledgements}


\end{document}